\documentclass{pasj01}
\draft
\begin{document}
\Received{}
\Accepted{}
\title{X-ray spectra of Sgr A East and diffuse X-ray background near the Galactic center}
\author{
Akiko \textsc{Ono}\altaffilmark{1},
Hideki \textsc{Uchiyama}\altaffilmark{2 ${\ast}$},
Shigeo \textsc{Yamauchi}\altaffilmark{1},
Masayoshi \textsc{Nobukawa}\altaffilmark{3},
Kumiko K. \textsc{Nobukawa}\altaffilmark{1},
and 
Katsuji \textsc{Koyama}\altaffilmark{4}
}
\altaffiltext{1}{Department of Physics, Nara Women's University, Kitauoyanishimachi, Nara 630-8506}
\altaffiltext{2}{Faculty of Education, Shizuoka University, 836 Ohya, Suruga-ku, Shizuoka 422-8529}
\email{uchiyama.hideki@shizuoka.ac.jp}
\altaffiltext{3}{Faculty of Education, Nara University of Education, Takabatake-cho, Nara 630-8528}
\altaffiltext{4}{Department of Physics, Graduate School of Science, Kyoto University, \\
Kitashirakawa-oiwake-cho, Sakyo-ku, Kyoto 606-8502}

\maketitle

\begin{abstract}

This paper reports the analysis procedure and results of  simultaneous spectral fits of the Suzaku archive data for Sagittarius (Sgr) A East and the nearby Galactic center X-ray emission (GCXE). The results are that the mixed-morphology supernova remnant Sgr A East has a recombining plasma (RP) with Cr and Mn He$\alpha$ lines, and a power-law component (PL) with an Fe \emissiontype{I} K$\alpha$ line.  The nearby GCXE has a $\sim$1.5-times larger surface brightness than the mean GCXE  far from Sgr A East, although  the spectral shape is almost identical.  Based on these results, we interpret that the origins of the RP and the PL with the Fe \emissiontype{I} K$\alpha$ line are past big flares of Sgr A$^\star$.

\end{abstract}

\section{Introduction}

Sagittarius (Sgr) A East  is a non-thermal radio shell,  a radio  supernova remnant (SNR) \citep{Ekers1983}. 
In the shell, there are other objects, the compact non-thermal radio source (Sgr A$^\star$), central massive black hole,  spiral-shaped thermal gas streams and central star cluster (Sgr A West). These sources may  closely couple with Sgr A East, hence the origin of Sgr A East has been extensively studied.
In the X-ray band, the Sgr A East spectrum is composed of 2-temperature collisional ionization equilibrium (CIE) plasmas, with $\sim$1 keV and $\sim$5--7 keV \citep{Sakano2004, Park2005, Koyama2007b}. 
The abundances are larger than the solar and heavy elements are spatially concentrated in the center. 
The mass of the progenitor star is estimated to be 13--20$\,M_\odot$ (e.g.,  \cite{Maeda2002}).  
The age is $\sim10^{3}$--$10^{4}$ years, and hence a young-middle aged SNR of a core-collapsed supernova (CC SN). 
It is classified as a mixed-morphology SNR \citep{Rho1998}.  
In addition to the two thermal plasmas, a non-thermal power-law component (PL) is found in the Suzaku spectrum \citep{Koyama2007b}. 

The major X-ray background  near  Sgr A East is the Galactic center X-ray emission (GCXE) \citep{Koyama2007b, Koyama2018} , which is composed with  the low temperature plasma (LTP) in the $\sim$2--5 keV band, high temperature plasma (HTP) and PL associated by Fe \emissiontype{I} K$\alpha$ line in the $\sim$5--10 keV band . 
The mean spectrum of the latter two components  has the equivalent  width (EW) of the Fe \emissiontype{I} K$\alpha$ (at 6.4 keV) and Fe  \emissiontype{XXV} He$\alpha$ (at 6.7 keV)  lines (hereafter EW6.4 and EW6.7) of   $\sim$180 eV and  $\sim$510 eV, respectively. 
The global spectrum is given by a  $\sim$14 keV temperature plasma with the Fe abundance of $\sim$1.2 solar \citep{Nobukawa2016}, or the latter components of a $\sim$7.4 keV plasma with the Fe abundance of $\sim$1.25 solar associated with a PL of photon index $\sim$2  together with the Fe \emissiontype{I} K$\alpha$ line \citep{Uchiyama2013}.
To avoid confusion of the GCXE background, we define the terminologies; ``the nearby GCXE'' is the GCXE in the 3/4 ring (the white dashed line in figure 1), and ``the background GCXE'' is the GCXE in the same area of Sgr A East (the white solid circle in figure 1). 
Thus, real background for Sgr A East is ``the background GCXE''. 
However, the spectrum and flux of the background GCXE has not been yet determined by Chandra \citep{Park2005}, XMM \citep{Sakano2004} or Suzaku \citep{Koyama2018}. 

The spectrum of  the nearby  GCXE reported by XMM-Newton and Chandra is significantly different from the mean GCXE; typically, the EW6.4 and EW6.7 are $\sim$220 eV and $\sim$730 eV, respectively \citep{Heard2013}, or the Fe abundance is $\sim$0.7 solar  \citep{Muno2004}. 
Also, the nearby GCXE has a larger surface brightness \citep{Uchiyama2013, Heard2013} than the mean GCXE. 
Therefore, proper estimation of the nearby GCXE is essential to determine a reliable spectrum of Sgr A East. 
For this requirement, we apply a new analysis procedure of simultaneous fit with Sgr A East and the nearby GCXE, paying particular attention to the energy band of the Fe K-shell complex. 

In spite of well studied HTP plasma, 
the LTP in the GCXE is less certain than the HTP (e.g., \cite{Yamauchi2018}). 
We therefore ignore the spectrum in the energy band below 2.36 keV and focus on the HTP of the 5--10 keV band. The energy band of 2.36--5 keV, which includes the key line S \emissiontype{XVI} He$\alpha$ \citep{Uchiyama2013}, is used for the qualitative estimate of the contribution of the LTP to the 5--10 keV band. 
In this paper, the distance to Sgr A$^\star$ is 8 kpc (e.g., \cite{Reid1993, Gillessen2009}), and quoted errors are in the 90 \% confidence limits. 

\section{Observations} 

Survey observations in the Galactic Center region were carried out with the X-ray Imaging Spectrometer (XIS: \cite{Koyama2007a}) onboard the Suzaku satellite \citep{Mitsuda2007}. 
The XISs were composed of 4 CCD cameras placed on the focal planes of the thin foil X-ray Telescopes (XRT: \cite{Serlemitsos2007}).  XIS\,1 employed a back-side illuminated (BI) CCD, while XIS\,0, 2, and 3 have front-side illuminated (FI) CCDs.  The field of view (FOV) of the XIS was \timeform{17'.8}$\times$\timeform{17'.8}.
The observation log is listed in table 1.

Since the spectral resolution of the XIS was degraded due to the radiation of cosmic particles, 
the spaced-row charge injection (SCI) technique was applied to restore the XIS performance \citep{Uchiyama2009}.  
The effective observation time of Sgr A East is $\sim$240 ks, which is far longer  than that in the previous Suzaku report \citep{Koyama2007b}.


\section{Analysis and Results} 

\subsection{Data Reduction} 

The XIS data in the South Atlantic Anomaly, during the earth occultation, and at the low elevation angle from the earth rim of $<5^{\circ}$ (night earth) and $<20^{\circ}$ (day earth) are excluded.  Removing hot and flickering pixels, the data of the Grade 0, 2, 3, 4, and 6 are used. The XIS pulse-height data are converted to Pulse Invariant (PI) channels using the {\tt xispi} software in the HEAsoft 6.19,  and the calibration database version 2016-06-07.
Figure 1 is the X-ray image of the 6.55--6.80 keV band in the area of Sgr A East and the nearby GCXE, where the non X-ray background (NXB) by {\tt xisnxbgen} \citep{Tawa2008} is subtracted. 
The color image and the green solid contours show the Suzaku image. 
These are similar to the smoothed Chandra image and the XMM image (see, figure 2e of \cite{Maeda2002}, and figure 1c of \cite{Sakano2004}). 
The spectra are made from  the areas of the  $\timeform{1'. 6}$ radius  circle (hereafter, the Sgr A East area) and  the 3/4 ring  of $\timeform{3'.0}$ -- $\timeform{5'.0}$ radius around Sgr A$^\star$ excluding  eastern bright X-ray reflection nebula (XRN) regions (\cite{Park2004, Koyama2018}, references therein) (hereafter, the nearby GCXE area), respectively (see figure 1).

\begin{table*}[t] 
\caption{List of data used for spectral analyses.}
\begin{center}
\begin{tabular}{lcccccc} 
\hline 
Observation ID 	&\multicolumn{2}{c}{Pointing position} 	&\multicolumn{2}{c}{Observation time (UT)}	& Exposure time$^{\ast}$	&Spectrum$^{\dagger}$\\
 					& $l(^\circ)$	& $b(^\circ)$				& Start						&End		& (ks)	&	\\
\hline
100027010		&0.057		&$-0.074$ 						&2005-09-23 07:18:25	&2005-09-24 11:05:19	&44.8	&Sgr A East / GCXE\\
100037040		&0.057		&$-0.074$						&2005-09-30 07:43:01	&2005-10-01 06:21:24	&42.9	&Sgr A East / GCXE\\
100048010		&0.057		&$-0.074$						&2006-09-08 02:23:24	&2006-09-09 09:06:15	&63.0	&Sgr A East / GCXE\\
100027020		&$-0.247$		&$-0.046$						&2005-09-24 14:17:17	&2005-09-25 17:27:19	&42.8	&Sgr A East / GCXE\\
100037010		&$-0.247$		&$-0.046$						&2005-09-29 04:35:41	&2005-09-30 04:29:19	&43.7	&Sgr A East / GCXE\\
501008010		&$-0.154$		&$-0.191$						&2006-09-26 14:18:16	&2006-09-29 21:25:14	&129.6	&GCXE\\
501009010		&$-0.074$		&0.178						&2006-09-29 21:26:07	&2006-10-01 06:55:19	&51.2	&GCXE\\

\hline 
\end{tabular}
\end{center}
$^{\ast}$ Effective exposure time after screening described in text. \\
$^{\dagger}$ This column shows whether the data were used to make Sgr A East or the nearby GCXE spectrum.\\
\end{table*}

\begin{figure} 
  \begin{center}
    \includegraphics[width=12cm]{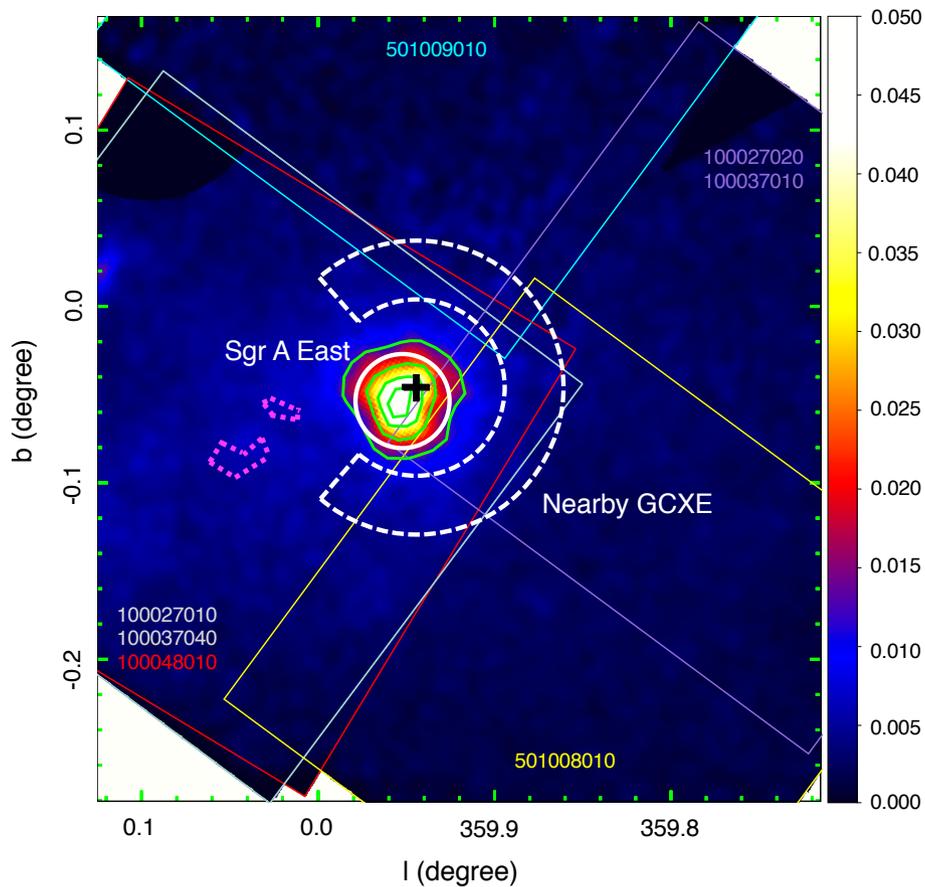}
\end{center}
\caption{The 6.55--6.80 keV band (Fe K-shell line complex)  image of XIS in the Galactic coordinate. 
The color image and the green solid contours show the Suzaku image. 
The color bar is X-ray surface brightness of liner scale in arbitrary unit.
The brightest region shown by the white solid circle is the Sgr A East area, while the 3/4 ring with the white dashed line is the nearby GCXE area.  
The black cross shows the position of Sgr A$^{\star}$.
The magenta dotted lines show the regions of  bright XRNe found by Chandra (\cite{Park2004}; filament 1 and 2). 
The nearby GCXE and Sgr A East are out of the XRN regions.
The borders of the field of view (FOV) of each observations (table 1) are separately shown by the gray, red, purple, cyan, and yellow lines, where those of ID100027010 and ID100037040, (also ID100027020 and ID100037010) are overlapped. 
}

\label{fig:sample}
\end{figure}

\subsection{Model Spectra of the Nearby GCXE and Sgr A East} 

Our main objective is spectral analysis in the Fe K-shell band (the $\sim$5--10 keV band),  both for the nearby GCXE and Sgr A East.  Accordingly, we ignore the energy band below 2.36 keV in the following spectral analysis.
The GCXE has been known to be composed of several different class of sources (e.g., \cite{Koyama2018}), and the spectrum has been successfully modeled by  high and low temperature CIE plasmas (HTP and LTP), with linked abundances for the both plasma\footnote{Since the GCXE is composed of many classes of sources with different temperatures and abundances \citep{Koyama2018}, assuming an average value of the estimated abundances for each element for both spectral components would be reasonable.}. In addition, a PL with Fe \emissiontype{I} K lines is included.
In the analysis of both the nearby GCXE and Sgr A East spectra, the cosmic X-ray background (CXB) compiled by \citet{Kushino2002} is included.

The model spectrum of Sgr A East is assumed to be a combination of ejecta and interstellar medium (ISM). 
We try two fits with the ISM models which are either a $\sim$ solar abundances plasma, or a plasma with same abundances as the those of the diffuse plasma of the nearby GCXE. 
No essential difference between these two fits is found. 
This paper refers to the results of the latter case, because Sgr A East is located in the GCXE, which would be largely affected by Sgr A$^{\star}$ activity, high star burst activity, strong magnetic field and etc,  and hence usual ISM with $\sim$solar abundance may not be applied.

\subsection{Simultaneous Fit of Sgr A East with the Nearby GCXE} 
The process of simultaneous-fit is analogous to solving simultaneous equations of $aX+bY=A$, and $ cX+dY = B$,  where $X$ and $Y$ are model functions of Sgr A East and GCXE, and $A$ and $B$ are observed spectra from the  Sgr A  East area and the nearby GCXE area.
The constant parameters of $a, b, c,$ and $d$ correspond to effective areas given by ARFs\footnote{Ancillary Response File (ARF) describes energy-dependent effective area which is calculated from the spatial distribution of the source and the photon accumulation region on the detector under the XRT response and contamination of the optical blocking filter \citep{Ishisaki2007}.}, ARF-1, 2, 3 and 4.
ARF-1 and ARF-2 are made using the Chandra image of Sgr A East \citep{Maeda2002, Park2005}. ARF-3 and ARF-4 are made from the Fe \emissiontype{XXV} He$\alpha$ distribution within $\timeform  {5'}$ from Sgr A$^\star$ \citep{Heard2013} of  a 2-exponential function with e-folding  longitude and latitude of $\timeform  {11'}$ and $\timeform  {9'}$, respectively. 
The fluxes of Sgr A East and the GCXE in the Sgr A East area are determined by ARF-1 and ARF-3 (parameters $a$, and $b$), while those in the nearby GCXE area are determined by ARF-2 and ARF-4 (parameters $c$, and $d$), respectively. 
The spectral parameters and fluxes of $X$ and $Y$ depend on both of $A$ and $B$. 
Therefore, all the relevant parameters are coupled with each other complicatedly.
The simultaneous fit with multiple ARFs can determine these parameters separately by $\chi^2$ minimizing process, which is  essential point of this fitting method.


The GCXE background by many previous authors (e.g., \cite{Sakano2004, Maeda2002}) were taken from the region around the nearby GCXE, not from the background GCXE of Sgr A East. 
Hence, their GCXE backgrounds were always underestimated, because the surface brightness of the GCXE near Sgr A East ($\lesssim \timeform{10'}$ radius) shows significant increase toward Sgr A$^\star$  \citep{Heard2013}. In fact, the surface brightness of the nearby GCXE and the background GCXE are estimated to be $2.4\times10^{-13}$ and $3.5\times10^{-13}$
erg s$^{-1}$ cm$^{-2}$ arcmin$^{-2}$ (5--10 keV), respectively.
In our simultaneous fit, the ARF takes account of the spatially distribution of the GCXE. Thus, the spectra and fluxes of the background GCXE and the nearby GCXE should be reliably estimated, in particular in the high energy band (5--10 keV) features of Fe peaked elements.  
As is shown in figure 1, each pointing position (Observation ID) of table 1 covers only a fraction of the nearby GCXE area. Therefore, the unit of the vertical axis (counts s$^{-1}$ keV$^{-1}$) in figure 2c is not proportional to  the flux of the full nearby GCXE area (the 3/4 ring). The effective area is calculated using ARFs generated with {\tt xissimarfgen} \citep{Ishisaki2007}, and is $\sim$45\% of the full 3/4 ring area. 

At first, we assume that Sgr A East spectrum is a composite of two CIE plasma, which represent the ejecta and ISM plasma \citep{Sakano2004, Park2005, Koyama2007b}. 
This two CIE model assumes $\sim$1 solar abundances of Mn and Cr, which is a same model as those of commonly used and accepted previously (e.g., \cite{Sakano2004, Park2005}).   
The simultaneous fit reveals that the two CIE model shows clear excess at the energies of Mn and Cr, indicating that Mn and Cr are over abundant ($>$ 1 solar).
We also find excess at 6.4 keV of Fe \emissiontype{I} K$\alpha$ line (see figure 2a). 
Adding a PL plus Fe \emissiontype{I} K component with free parameter of equivalent width, a better fit with $\chi^2$/d.o.f. of 286/228=1.25 (null probability is 0.6 \%.) is obtained.  
Hereafter, we named this model as the conventional 2-CIE model (in short, 2-CIE). 
The best-fit 2-CIE model and parameters are given in figure 2a and the $2^{nd}$, $3^{rd}$ columns in table 2, respectively. 
In figure 2a, we find significant data excesses from the 2-CIE model in the energy band of 5--10 keV, where the radiative recombination continuum (Fe \emissiontype{XXV} RRC), Cr \emissiontype{XXIII} He$\alpha$ and Mn \emissiontype{XXIV} He$\alpha$ exist. 
The excess of the Fe \emissiontype{XXV} RRC is direct evidence for the recombining plasma (RP) (e.g., \cite {Ohnishi2014}), because the relevant plasma should include significant fraction of Fe  \emissiontype{XXVI} ions, more than that of the CIE plasma.

In order to check the significance of the excesses at the Fe-RRC, Mn and Cr line energies in the 2-CIE model, we apply two RP model (for the residual at Fe-RRC) including Mn and Cr abundances as free parameters (for the residuals at Mn and Cr lines). This two RP model includes the PL with Fe \emissiontype{I} K lines  (hear and after, 2-RP model).
Then, most of the residuals disappear (see figure 2b) with statistically highly acceptable results of $\chi^2$/d.o.f.=241/224=1.08. 
The reduction of the $\chi^2$ value from the 2-CIE to the 2-RP is 45, which is equally shared to the Fe-RRC and Mn, Cr residuals.  
If we limited the energy band of 5--10 keV, where Fe-RRC and Cr, Mn lines exist, the $\chi^2$/d.o.f. of the 2-CIE and 2-RP models are 142/90=1.58 and 100/86=1.16, respectively.  This large difference also justify to apply the 2-RP model instead of 2-CIE. 
We further apply the F-test judgment and find significant improvement from the 2-CIE fit ($\chi^2$/d.o.f.=286/228=1.25) to the 2-RP fit ($\chi^2$/d.o.f.=241/224=1.08) with  more than 99.99 \% of null probability. 
The best-fit Sgr A East and the nearby GCXE spectrum  in the case of the 2-RP model  are given in figure 2b and 2c, respectively. The best-fit  parameters are shown in the $4^{th}$ and $5^{th}$ columns of table 2. 



\begin{table*} 
\caption{The best-fit parameters for Sgr A East and the nearby GCXE in the cases of the 2-CIE  or 2-RP model for Sgr A East.}
\label{tab:first}
\begin{center}
 \begin{tabular}{lccccc}
      \hline
      &\multicolumn{2}{c}{In the case of 2-CIE model for Sgr A East}&&\multicolumn{2}{c}{In the case of 2-RP model for Sgr A East}\\
							& Sgr A East & nearby GCXE &~~~~~&Sgr A East  & nearby GCXE 	\\
\hline
							&ISM (CIE)				&LTP (CIE) &~~~~~&ISM (RP)	&LTP (CIE) \\
$kT_{\rm e}^{*}$    				& $1.05\pm0.05$		& $1.04\pm0.03$ &~~~~~& $0.95\pm0.10$	& $1.05\pm0.03$\\
$kT_{\rm i}^{*}$					& -			& 		-	 &~~~~~& 10 (fixed)& 		-	\\
$Z_{\rm S}$$^{\dagger}$ 			& link to nearby GCXE	& link to HTP &~~~~~& link to nearby GCXE& link to HTP \\
$Z_{\rm Ar}$ $^{\dagger}$ 		& link to nearby GCXE	& link to HTP &~~~~~& link to nearby GCXE& link to HTP \\
$Z_{\rm Ca}$$^{\dagger}$ 		& link to nearby GCXE	& link to HTP &~~~~~& link to nearby GCXE& link to HTP \\
$n_{\rm e}t^{\|}$ 				& -	&-				&~~~~~& $ >17$	&-\\
Norm.$^{\#}$ 					& $0.19\pm{0.03}$ 		& $0.35\pm{0.06}$  &~~~~~& $0.19\pm{0.06}$ 		& $0.33\pm{0.03}$\\
\hline
							&Ejecta (CIE)				&HTP (CIE) &~~~~~& Ejecta (RP) &HTP (CIE) \\
$kT_{\rm e}^{*}$				& $4.5\pm{0.1}$ 		& $7.4\pm$0.2		&~~~~~& $2.3\pm{0.2}$ 		& $7.4\pm$0.2	\\
$kT_{\rm i}^{*}$					& -			& 		-	 	&~~~~~& 10 (fixed)			& 		-	 \\
$Z_{\rm S}$$^{\dagger}$			&   $0.0 (<2.0)$		& $1.7\pm0.1$ 		 &~~~~~& $0.3 (<1.4) $		& $1.7\pm0.1$ \\
$Z_{\rm Ar}$$^{\dagger}$			& $0.8 (<2.4)$			& $1.3\pm0.1$		&~~~~~& $0.4 (<1.0)$	& $1.4\pm0.1$\\
$Z_{\rm Ca}$$^{\dagger}$			& $0.7 (<1.8)$ 		& $1.5\pm0.1$		&~~~~~& $1.0 \pm0.5$ & $1.6\pm0.1$\\
$Z_{\rm Cr}$$^{\dagger}$			& 1.0 (fixed) 			& 1.0 (fixed) 		 &~~~~~&$4.4\pm1.6$	& 1.0 (fixed) 		\\
$Z_{\rm Mn}$$^{\dagger}$ 		& 1.0 (fixed) 	  		& 1.0 (fixed) 		&~~~~~& $14\pm8$  			& 1.0 (fixed) 	\\
$Z_{\rm Fe}$$^{\dagger}$ 			 & $1.8\pm0.4$	  		& $1.25\pm0.05$	 &~~~~~&$1.5\pm0.3$	  		& $1.25\pm0.05$\\
$Z_{\rm Ni}$$^{\dagger}$			& $0.4 (<1.3)$			& 1.25 (link to Fe) 	&~~~~~&$1.5 (<3.0)$			& 1.25 (link to Fe) \\
$n_{\rm e}t^{\|}$ 			    	& -	 	& -					&~~~~~&	$6.3\pm{0.5}$	 & -	\\
Norm.$^{\#}$ 					& $0.013\pm{0.009}$ 	& $0.041\pm{0.001}$   &~~~~~& $0.034\pm{0.017}$ 	& $0.041\pm{0.001}$\\
\hline
							&\multicolumn{2}{c}{Power law + Fe \emissiontype{I} K$\alpha$ }&~~~~~&\multicolumn{2}{c}{Power law + Fe \emissiontype{I} K$\alpha$ }\\
$\Gamma$ &					$1.0\pm0.9$&  $1.7\pm0.1$  &~~~~~& $1.0\pm0.8$&  $1.7\pm0.1$ \\
PL norm.$^{**}$ 				& $0.40\pm0.27$		&	$1.95\pm0.07$	&~~~~~&$0.45\pm0.22$		&	$1.96\pm0.07$\\	
EW6.4 (eV) 					&  $200\pm{140}$	& 460 (fixed)$^{\dagger\dagger}$  &~~~~~&  $160\pm{80}$	& 460 (fixed)$^{\dagger\dagger}$\\
\hline
$N_{\rm H}$$^{\|\|}$ 				&$15\pm$1 &$8.9\pm0.2$ &~~~~~&$15\pm1$ &$8.7\pm0.2$ \\
\hline
$\chi^2$/d.o.f.  					&\multicolumn{2}{c}{286/228=1.25}  &~~~~~& \multicolumn{2}{c}{241/224=1.08}   \\      
   \hline
    \end{tabular}
\end{center}
$^*$ Units are keV. $kT_{\rm e}$ is the electron temperature. $kT_{\rm i}$ is the initial ionization temperature at $n_{\rm e}t=0$.  
$^{\dagger}$ Abundances relative to the solar value of \citet{Anders1989}.
$^{\|}$  Unit is $10^{11}$ s cm$^{-3}$. 
$^{\#}$ Defined as 10$^{-14}$$\times$$\int n_{\rm H} n_{\rm e} dV$ / (4$\pi D^2$) (cm$^{-5}$),
where $n_{\rm H}$, $n_{\rm e}$, and $D$  are   hydrogen density (cm$^{-3}$),
electron density (cm$^{-3}$) and distance to Sgr A East (cm), respectively. 
$^{**}$ Unit is photons s$^{-1}$ cm$^{-2}$ keV$^{-1}$ at 6.4 keV. 
$^{\dagger\dagger}$  Fixed to the value of the mean GCXE \citep{Uchiyama2013}.
$^{\|\|}$ The hydrogen column density of the interstellar absorption in unit of $10^{22}$ cm$^{-2}$.
\end{table*}

\begin{figure}[htb] 
  \begin{center}
      \includegraphics[width=7.5cm]{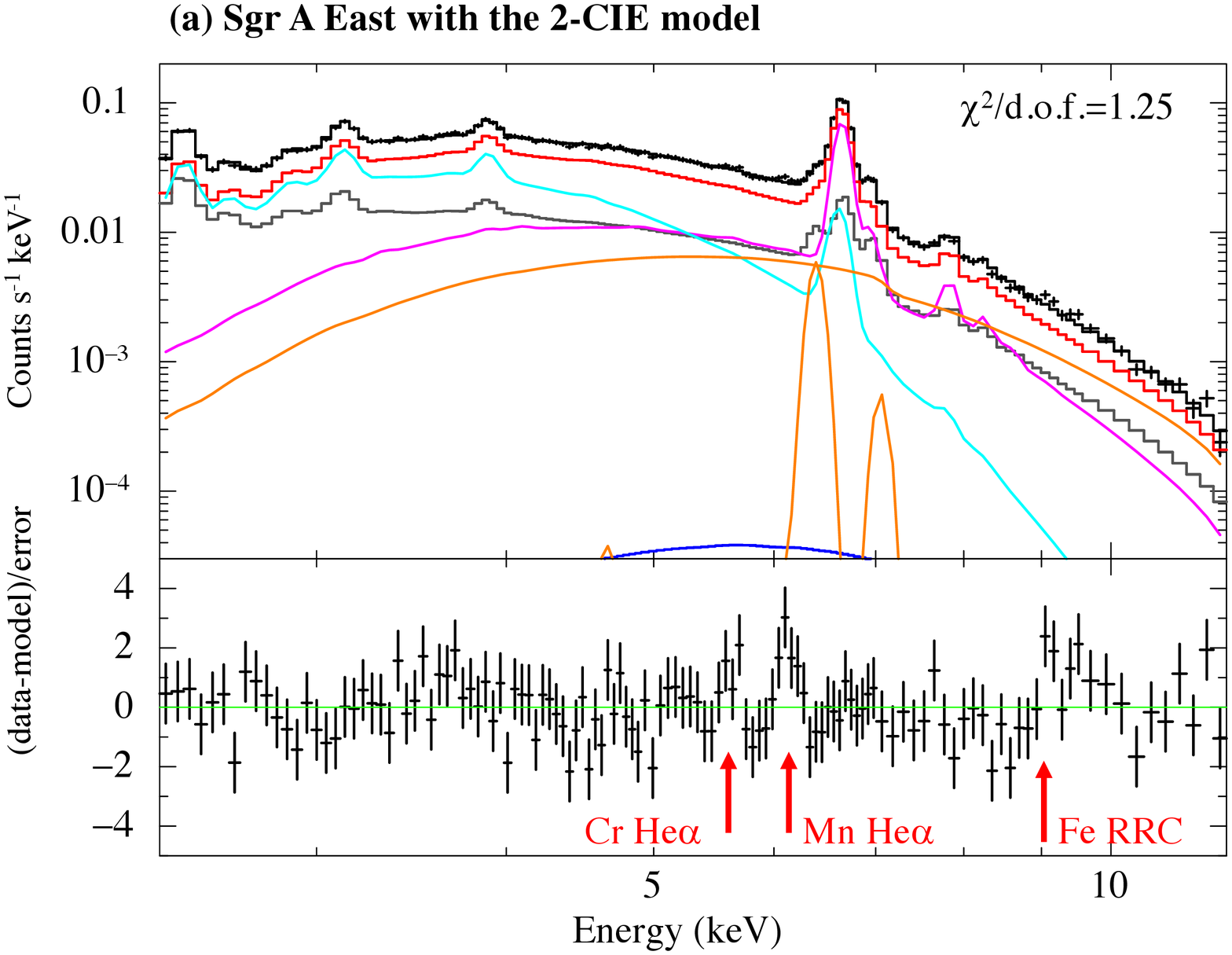}\\ \vspace{3mm}
       \includegraphics[width=7.5cm]{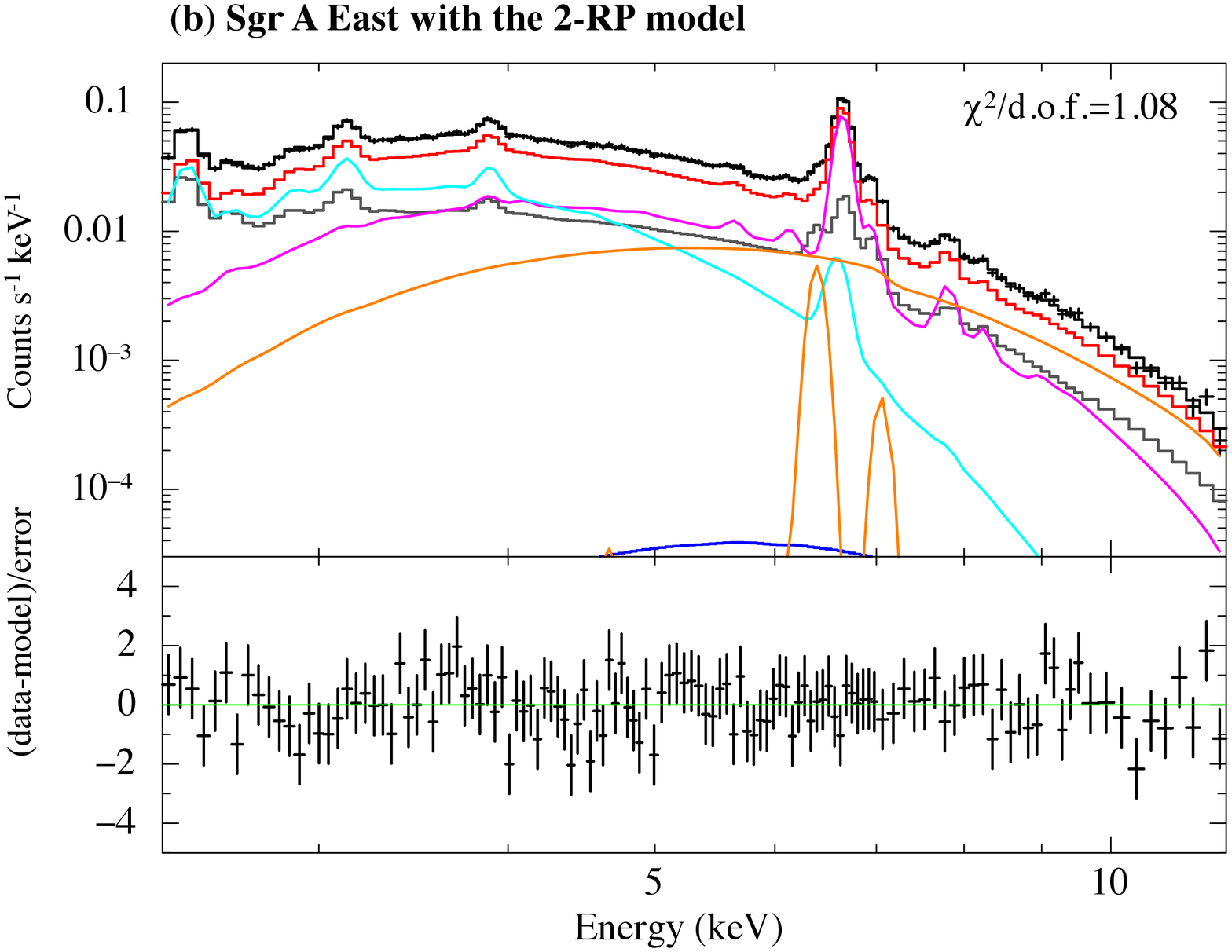} \\\vspace{3mm}
       \includegraphics[width=7.5cm]{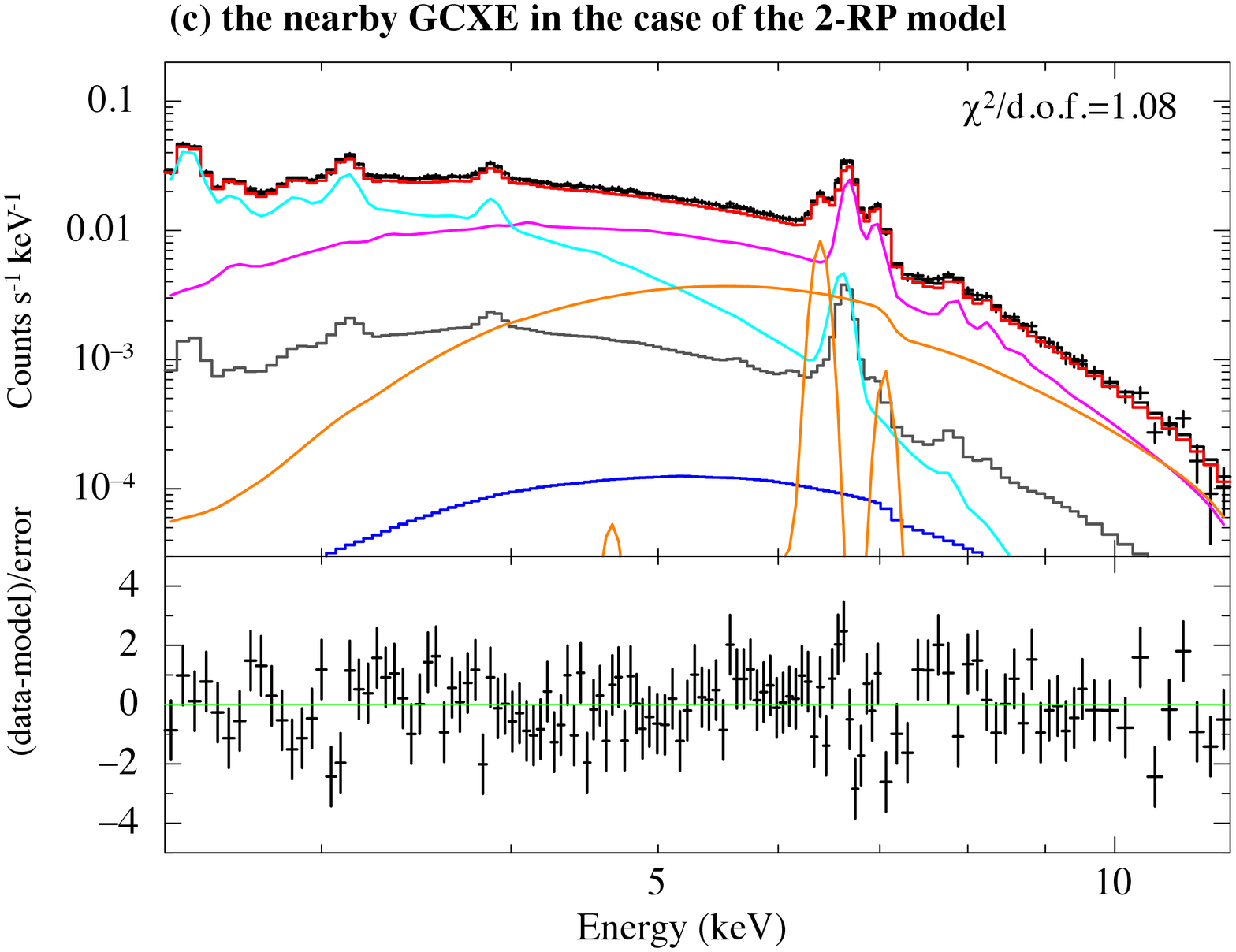} 
  \end{center}
\caption{(a,b) The best-fit Sgr A East models of  2-CIE  and 2-RP.
 The magenta, cyan,  orange, blue and gray lines show the ejecta plasma, ISM plasma, PL+Fe \emissiontype{I} K lines, CXB and ``the background GCXE'', respectively.  The red line shows the total model of Sgr A East SNR (ejecta+ISM+PL+Fe \emissiontype{I} K lines). The black line shows the sum of all the model components (Sgr A East SNR+the background GCXE+CXB).
 (c) The best-fit nearby GCXE model in the case of  2-RP. The magenta, cyan,  orange, blue and gray lines show the HTP, LTP,  PL+Fe \emissiontype{I} K lines, CXB and the contamination of Sgr A East, respectively. The red line shows the total model of the nearby GCXE (HTP+LTP+PL+Fe \emissiontype{I} K lines). The black line shows the sum of all the model components (the nearby GCXE+the contamination of Sgr A East+CXB).}
\label{fig:sample}
\end{figure}

\section{Discussion} 

\subsection{X-Ray Spectrum of the Nearby GCXE} 
 
We find that the nearby GCXE has almost the same temperatures and abundances  as those of the mean GCXE (see table 2 and \cite{Koyama2018}). 
These results seem inconsistent with the results of Chandra and XMM-Newton \citep{Muno2004, Heard2013}.  
The reason of this apparent inconsistency is simple;  \citet{Muno2004} and \citet{Heard2013} reported that the significant contribution from a PL component was not required to describe the observed X-ray spectrum, while our GCXE has the PL with $~46\sigma$ significance (see table 2). 
The nearby GCXE has a high temperature of $\sim$7.4 keV with the Fe abundance of $\sim$1.25 solar, 
essentially the same spectrum as, but $\sim$1.5 times larger surface brightness than the mean GCXE.

\subsection{Contribution of Sgr A$^\star$ to Sgr A East}
The background GCXE is partially composed of many point sources including Sgr A$^\star$. 
The brightest point source Sgr A$^\star$ has the absorption-corrected luminosity of $2.4\times10^{33}$ erg s$^{-1}$ (2--10 keV; \cite{Baganoff2003}). 
Using the spectrum parameters of Sgr A$^\star$ ($\Gamma$=2.7, $N_{\rm H}=10^{23}$ cm$^{-2}$; \cite{Baganoff2003}), it is converted to the absorbed luminosity of $5.8 \times10^{32}$ erg s$^{-1}$ in the 5--10 keV band, which covers the essential energy band of Fe-K lines and RRC of Sgr A East.  
Sgr A$^\star$ exhibits weak flares of $<10^{35}$ erg s$^{-1}$ (2--10 keV) with the rate of $\sim$1 day$^{-1}$ and the flare duration is less than $\sim$ a few ks \citep{Neilsen2013}. 
The time averaged luminosity is estimated to be in the order of  $<10^{33}$ erg s$^{-1}$ (5--10 keV) using $\Gamma$ =2.3 \citep{Ponti2017}.
The observed luminosity of Sgr A East is $\sim7\times10^{34}$ erg s$^{-1}$ in the 5--10 keV band \citep{Koyama2007b}. 
Thus, the contribution of both of the stable (non-flare) and time-averaged flare fluxes of Sgr A$^\star$  is estimated to be at most $\sim$3 \%  of the Sgr A East  flux.
Although the nominal spatial resolution of XIS is $\sim1\arcmin$, the point spread function has a sharp cusp structure \citep{Serlemitsos2007}, which makes it possible to show a peak position of Sgr A$^\star$ flare with better than $\timeform{20''}$ accuracy \citep{Uchiyama2008}. The  Suzaku image  shows no peak nor large enhancement at the position of Sgr A$^\star$ (see figure 1).  
Accordingly, the effects of the Sgr A$^\star$  flux to the spectrum of Sgr A East are negligible. 
 
\subsection{X-Ray Spectrum of Sgr A East} 
We find that the 2-RP model gives better fit than the 2-CIE model to the updated Suzaku spectrum (see figure 2).  
In addition to the better $\chi^2$ value of the 2-RP model than the 2-CIE model, we preferred the 2-RP model than the 2-CIE model, because the former  predicts the presence of the RP, and the He$\alpha$ lines of rare elements, Cr and Mn, and the Fe \emissiontype{I} K$\alpha$ line in the ejecta plasma for the first time. 

The temperature of $\sim$4.5 keV in the 2-CIE fit is high even for the young SNR. In fact, the youngest Galactic CC-SNR Cas A has only $\sim$ 2--4 keV \citep{Maeda2009}, or has a steep power-law index of $\sim$3 \citep{Sato2017}. This high temperature plasma in the 2-CIE fit is reduced to $\sim$2.3 keV in the 2-RP fit (the $2^{nd}$ and $4^{th}$ columns of table 2), which is more typical temperature for a young-middle aged CC SNR (e.g., Cas A, \cite{Maeda2009}). Even this temperature produces Fe \emissiontype{XXVI} Ly$\alpha$ line, which is another important aspect of the 2-RP model. The abundance pattern of S, Ar, Ca, Fe and Ni of ejecta favors a CC SN origin of the low mass side  \citep{Maeda2002}.
One may argue that Fe abundance in the ejecta of SNR, is not enhanced (table 2). 
This seems to be inconsistent with the previous CIE fit. 
In fact, \citet{Park2005} found that Fe abundance is 5.8$^{+1.7}_{-1.1}$ solar at the center. 
On the other hand, our 2-RP fit gives the average Fe abundance to be $1.5\pm{0.3}$ solar. 
This apparent inconsistency is not surprising because the previous work is CIE fit for the Sgr A East spectra with under-estimated GCXE background, but ours is 2-RP fit with the proper subtraction of GCXE background (see the 2$^{nd}$ paragraph of section 3.3).  
In addition, the Fe abundance of $\sim$5.8 solar is the result of the very small ($\sim \timeform{20"}$) Fe-rich central region  \citep{Park2005}.
\citet{Sakano2004} and \citet{Park2005} find Fe increases toward the central region of Sgr A East.  
Our result is the mean abundance of the whole ejecta ($\timeform{1'.6}$-radius region) and naturally smaller than that of the Fe-rich center.
We should note that the mean Fe abundance of $\sim$1.5 solar is not unreasonable for the ejecta of CC-SNRs.

Due to the low abundance of Mn, Mn \emissiontype{XXIV} He$\alpha$ lines have been detected from  only 6 SNRs \citep{Yang2013}. 
Among them, W49B is a unique SNR exhibiting Fe-RRC  (e.g., \cite{Ozawa2009}).  
In all of those 6 SNRs, the flux ratios of  Mn/Cr are within  standard models (e.g., \cite{Woosley1995, Sukhbold2016}).  
Sgr A East, however, has a far larger Mn/Cr ratio, which is out of the standard models. 
We speculate that Mn is over-produced in a neutron rich region, or in neutron star.  

In the mixed-morphology SNR, the current  scenario for the origin of the RP is either electron cooling by cold molecular clouds \citep{Kawasaki2002} or adiabatic expansion through the clouds  \citep {Masai1994} (rarefaction). The 2-RP model requires the initial ionization temperature ($kT_{\rm i}$) of $\sim10$ keV, which is not obtained by the conventional electron cooling scenario.  Therefore, we speculate other possibility, an X-ray photo-ionization of Sgr A$^\star$ flares, which also explain the presence of Fe \emissiontype{I} K$\alpha$ line.
In this scenario, the structure of RP is determined by the parameter of $\xi=L /(n_{\rm e} R^2$) \citep{Kallman2004}, where $n_{\rm e}$, $R$ and $L$ are the electron density of the ejecta plasma (cm$^{-3}$),  the distance (cm) from Sgr A$^\star$ and the luminosity (erg s$^{-1}$) of the past flare of Sgr A$^\star$, respectively. 
If log $\xi \gtrsim4$, the ejecta becomes H-like and naked Fe dominant plasma (RP).
From the volume emission measure of the ejecta, $n_{\rm e}$ is estimated to be $\sim4$ cm$^{-3}$.
Then,  with the best-fit $n_{\rm e}t$ of 6.3$\times10^{11}$ s cm$^{-3}$, log$\xi = 4$ and $R\sim$1 pc, the time ($t$) after the photo ionization and flare luminosity ($L$) are estimated to be $\sim10^4$ years and $\sim10^{42}$ erg s$^{-1}$, respectively.    
Thus, in Sgr A East of $R\sim$1 pc, the RP would be produced by Sgr A$^\star$ flares
of $\sim10^4$ years ago with the luminosity of $L\sim10^{42}$ erg s$^{-1}$.
The flare  luminosity is in between those of $\sim$100--800 years ago ($\sim10^{39}$ erg s$^{-1}$; \cite{Ryu2009}, 2013) and that of $\sim10^5$ years ago ($\sim10^{44}$ erg s$^{-1}$; \cite{Nakashima2013}).

The PL components are found in both Sgr A East and the nearby GCXE spectra (table 2) with the significance levels of 3.3$\sigma$ and 46$\sigma$, respectively. 
\citet{Muno2008} reports  12 non-thermal filaments in the Sgr A East area. 
We make the summed spectrum of the 12 filaments and found that the spectrum is a PL with the luminosity of $7.4\times10^{33}$ erg s$^{-1}$ (5--10 keV), which is only 24 \%  of the best-fit PL luminosity of Sgr A East.  In the summed spectrum, no significant Fe \emissiontype{I}  K$\alpha$ line is detected. 
Therefore a major fraction of the PL plus Fe \emissiontype{I}  K$\alpha$ line would be due to undetected components. 
One speculation is that there are many faint XRNe of small $N_{\rm H}$ cloudlets, which would be hard to detect with the present instruments.


\section{Conclusion}

We have performed simultaneous spectral fits for Sgr A East and the nearby GCXE.
The results are:      

\begin{itemize}

\item  RRC structure of Fe \emissiontype{XXV}, and K-shell lines of Cr \emissiontype{XXIII} He$\alpha$, Mn\emissiontype{XXIV} He$\alpha$ and Fe \emissiontype{I} K$\alpha$ are discovered from Sgr A East.

\item The spectrum of Sgr A East is nicely explained by two RP components.

\item  The origin of the RP is probably due to the past big flares of Sgr A$^\star$ with the luminosity of  $\sim10^{42}$ erg s$^{-1}$.

\item  The nearby GCXE has a similar spectrum to the mean GCXE,  although the surface brightness is $\sim$1.5 times larger than that of the mean GCXE.
 
\end{itemize}
   
\section*{Acknowledgement}

The authors are grateful to all members of the Suzaku team. 
This work was supported by the Japan Society for the Promotion of Science (JSPS) 
KAKENHI Grant Numbers JP25887028 (HU), JP15H02090, JP17K14289 (MN), JP24540232 (SY) and JP16J00548 (KKN) and Nara Women's University Intramural Grant for Project Research (SY).


\end{document}